# Brightening spin- and parity-forbidden interlayer excitons through rhombohedral stacking

Haodong Hu[1,2,*], Chih-En Hsu[3,4,*], Kenji Watanabe[5], Takashi Taniguchi[6], Hung-Chung Hsueh[4], Zhenglu Li[3,†], and Ziliang Ye[1,2,‡]

[1]Department of Physics and Astronomy, The University of British Columbia, Vancouver, BC V6T 1Z1, Canada

[2]Quantum Matter Institute, The University of British Columbia, Vancouver, BC V6T 1Z4, Canada

[3]Mork Family Department of Chemical Engineering and Materials Science, University of Southern California, Los Angeles, California 90089, USA

[4]Department of Physics, Tamkang University, New Taipei City 251301, Taiwan

[5]Research Center for Functional Materials, National Institute for Materials Science, 1-1 Namiki, Tsukuba 305-0044, Japan

[6]International Center for Materials Nanoarchitectonics, National Institute for Materials Science, 1-1 Namiki, Tsukuba 305-0044, Japan

[*]These authors contributed equally to this work.

[†]Contact author: zhenglul@usc.edu

[‡]Contact author: zlye@phas.ubc.ca

## Abstract:

Brightened symmetry-forbidden excitons can provide critical information about the electronic structure and many-body interactions in semiconductors. Recently, interlayer atomic registry has emerged as a new way to control atomically thin layered materials in both ground and excited states. Here we show that the rhombohedral stacking in a 3R-$MoS_2$ homobilayer can brighten two symmetry-forbidden interlayer excitons: one is a spin-unlike 1s exciton and the other is a parity-odd 2p exciton. The nature of these interlayer excitons is revealed by their g-factors as well as the hybridization patterns observed when they are tuned to resonate with intralayer excitons. Surprisingly, the hybridized 2p exciton exhibits a constructive quantum interference between the acquired transition dipoles, revealing a hidden gauge-invariant phase contrast between the constituent electronic states. Our results establish stacking control as a powerful approach for modulating quantum phases of excitons in 2D semiconductors.

## Main:

Among excitons in atomically thin transition metal dichalcogenides (TMDs), some are forbidden by symmetries such as spin and parity to respond to linear optical probes at normal incidence, known as dark excitons [1–3]. Albeit optically inactive, dark excitons serve as important reservoirs for optical excitations and can reveal materials' electronic structure and many-body interaction information [3–7]. Significant effort has been made to identify and brighten them, i.e. increasing their oscillator strengths by applying large electric or magnetic fields to break the symmetry [6–11]. Recently, Yu et al. proposed that some regions of TMD heterobilayers can brighten spin-forbidden interlayer excitons (IXs) [12], but such an effect has remained elusive in experiments due to the coexistence of multiple stacking configurations within each moiré unit cell [13,14]. Here we show that the uniform rhombohedral stacking in a 3R-$MoS_2$ homobilayer can brighten not only a spin-forbidden IX but also a parity-forbidden IX with an unusual quantum interference effect. At resonant conditions, these brightened IXs acquire oscillator strengths comparable to that of the intralayer excitons.

Interlayer excitons (IXs) exist in few-layer TMDs, with electrons and holes residing in different layers [15,16] and energies highly tunable through their large dipole moments [17,18]. In a rhombohedral-stacked TMD bilayer, the interlayer atomic registry uniformly breaks the mirror and inversion symmetries, leading to an effective type-II band alignment at the K valleys of the Brillouin zone with a layer-contrasting valley angular momentum [19–24], rendering certain symmetry-forbidden IXs weakly bright. Using an out-of-plane electric field, we can further brighten these IXs by tuning them to resonate with intralayer excitons [25–28]. Near the resonances, we observe the intra- and inter-layer excitons hybridize with rich hybridization patterns, through which we can reveal the nature of each brightened IX.

Experimentally, we probe these hybridizations in dual-gated 3R-$MoS_2$ bilayers using low-temperature reflection contrast spectroscopy. The bilayers are directly exfoliated from 3R bulk crystals with no preexisting domains, so the stacking configuration is stable throughout the measurements [23,29,30]. Fig. 1c shows the energy-derivative of the reflection contrast (dRC) spectra in a typical device at various electric fields ($E$). Two electric-field-independent optical resonances appear as vertical lines around 1.9 eV with a 12-meV splitting, corresponding to the intralayer A excitons in top and bottom layers ($X_T$ and $X_B$) [24]. Due to the local field effect, dRC spectra in this device resemble the imaginary part of the permittivity. Interestingly, near the electric fields of -0.05 V/nm and -0.1 V/nm, two new features emerge near $X_T$ and $X_B$. These emerging features disperse linearly with the field, intercepting intralayer excitons with clear hybridization gaps, and they are only visible near intralayer excitons. By fitting their energy-field slopes, we find they have out-of-plane dipole moments of 0.66 and 0.65 $e \cdot nm$, respectively. These evidence suggest that the dispersive features are associated with positive-dipole IXs ($IX_1$ & $IX_2$), whose energies increase with increasing electric field, corresponding to electrons and holes residing in the top and bottom layers, respectively. The IXs and their hybridizations are also observed in our first-principles GW-BSE calculations of the absorbance of a bilayer 3R-$MoS_2$ (Fig. 1d), except that the hybridizations appear at slightly different field and energy, likely due to the screening effect in physical devices.

Although the dispersive features can be qualitatively explained by the brightened IXs, there remain some inexplicable spectral contrasts. First, the two IXs exhibit distinct hybridization patterns: $IX_1$ has an avoided crossing with $X_B$ and crosses $X_T$ without opening any gap; $IX_2$ shows avoided crossings with both intralayer excitons, though the hybridization pattern is not a simple sum of two avoided crossings. Second, dRC spectra exhibit a pronounced asymmetric electric field dependence: no IX is observed with a negative slope in the positive field range. The asymmetry in field is too large to be accounted by the built-in interlayer potential only. Fundamentally, these spectral contrasts stem from the symmetry of the IX wavefunction, which dictates the hybridization selection rules with intralayer excitons.

With different rotational symmetries, excitons of different angular momenta do not hybridize [31]. The angular momentum of an exciton includes the spin and orbital parts of the constituent Bloch states with the exciton's envelope function [20,32–36], with the latter describing the correlations between electron and hole. The hybridization selection rule thus suggests a common spin-unlike exciton should not couple with a spin-like bright exciton. However, such hybridization is allowed in 3R-TMDs due to the additional layer-contrasting valley angular momentum introduced by the rhombohedral stacking, which we will explain in detail below. Finally, according to the GW-BSE theory, if the hybridization between two excitons is symmetry-allowed, the hybridization strength is determined by the wavefunction overlap between the underlying Bloch states, which is also critical to understanding our results.

To apply the hybridization selection rule in 3R bilayers, we first determine the effective angular-momentum quantum numbers (AMQNs) of the Bloch states at K valley of each layer. In monolayer TMDs, these AMQNs arise from the atomic orbitals and the Bloch phase of the bands, leading to the well-known valley selection rule [32–34]. In a 3R bilayer, the atomic registry between two layers further contributes to an additional layer-dependent Berry phase under $C_3$ rotation [12,19,24], giving rise to the layer-contrasting valley AMQN (Fig. 1b). Composed of a pair of electron and hole residing at band extrema with the same spin and trivial envelope function, the AMQN of a bright 1s exciton is equal to the AMQN difference between the two electronic states. As a result, the AMQN of IX is equal to -1 (2 modulo 3), opposite to that of $X_B$ or $X_T$ (+1). We label them with distinctive colors in Fig. 1b (blue versus red) to highlight that their hybridization is forbidden by symmetry. Such an IX therefore cannot account for the resonances observed in our experiments. Next, we show that $IX_1$ and $IX_2$ are the spin-unlike 1s and spin-like 2p IXs, respectively.

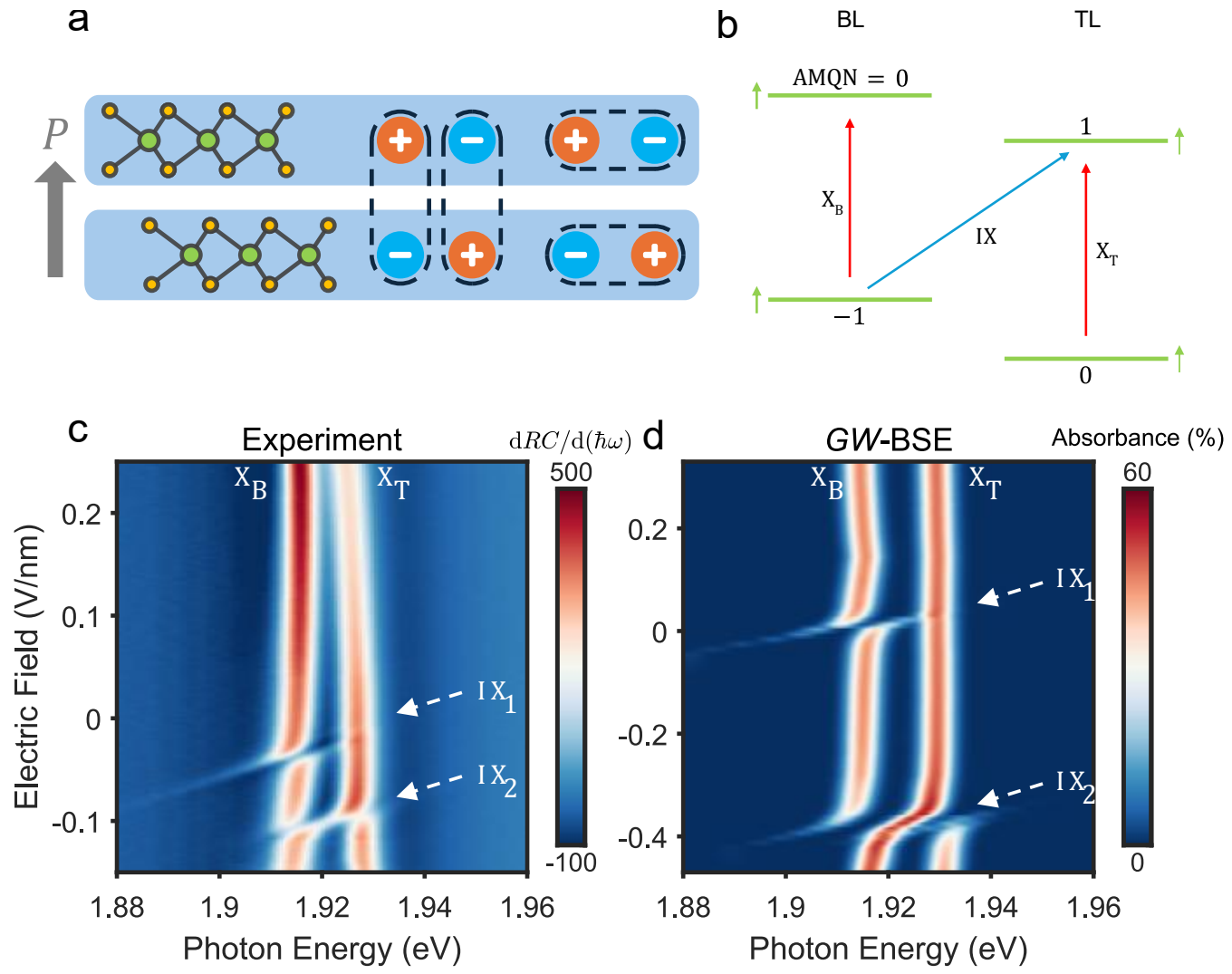

FIG. 1. Observation of two interlayer excitons in a bilayer 3R-$MoS_2$. (a) Schematic of a 3R-$MoS_2$ bilayer with an upward polarization P with all the possible exciton configurations. (b) Schematic of a type-II band alignment of band edges at the K valley between the top and bottom layers (TL & BL). A valley angular-momentum quantum number (AMQN) is assigned to each Bloch state. The exciton AMQN is equal to the AMQN difference between the electron and hole states, modulo three. Red and blue arrows denote the exciton with AMQN of +1 and -1

respectively. The spin AMQN is not included. (c) dRC spectra measured at 1.6 K as a function of the out-of-plane electric field. (d) Calculated GW-BSE absorbance [37,38].

*Spin-unlike 1s interlayer exciton*—Compared to the spin-like IX, the spin-unlike IX in bilayer 3R-MoS2 has a spin angular momentum that cancels the valley one, yielding a total AMQN identical to those of intralayer excitons (Fig. 2a). As a result, the spin-unlike IX can hybridize with the spin-like intralayer excitons. Here we attribute $\mathrm{IX_1}$ to the spin-unlike 1s exciton due to its asymmetric hybridization pattern. As discussed above, the exciton hybridization strength is determined by the wavefunction overlap of the constituent Bloch states. Between $\mathrm{X_T}$ and assigned spin-unlike 1s IX shown in Fig. 2a, they share neither the electron nor the hole band, and their hybridization is thus negligible. On the other hand, $\mathrm{X_B}$ and $\mathrm{IX_1}$ share the same valence band – their hole wavefunction overlap is significant. Consequently, the spin-unlike 1s IX hybridizes much more strongly with $\mathrm{X_B}$ than $\mathrm{X_T}$. We note that although $\mathrm{X_B}$ and $\mathrm{IX_1}$ do not share the same conduction band for the electron, they still have a finite wavefunction overlap due to Rashba-like spin-orbit-coupling-induced interband mixing [12,39,40].

Experimentally, $\mathrm{IX_1}$ indeed exhibits an avoided crossing with $\mathrm{X_B}$ and a plain energy-level crossing with $\mathrm{X_T}$ (Fig. 2b). To quantify the hybridization strengths, we fit the 2D dRC spectra with a three-level model including the couplings between $\mathrm{IX_1}$ and the two intralayer excitons (Supplementary information [41]). The fitted peak energies of the hybridized excitons are plotted against the electric field in Fig. 2c. The coupling strength between $\mathrm{IX_1}$ and $\mathrm{X_B}$, $t_{1B}$, is fitted to be 2.8 meV. In contrast, the coupling strength between $\mathrm{IX_1}$ and $\mathrm{X_T}$, $t_{1T}$, is below the noise floor. Furthermore, the asymmetric hybridization is also reflected in the relative oscillator strengths among three exciton branches. At the avoided crossing point, the first and second exciton branches have nearly identical oscillator strengths, about half of that of the third branch (Fig. 2d), meaning the spin-unlike IX acquires a considerable oscillator strength after hybridization, which could be used for efficient optical spin injection. By adjusting electric field, we can transfer the oscillator strength between the first and second branches while keeping the third constant. As we will show in Fig. 4, the spin-unlike configuration of $\mathrm{IX_1}$ is also confirmed by the g-factor measurement. Moreover, the spin-unlike nature of $\mathrm{IX_1}$ also explains the absence of its negative-dipole counterpart. As shown in Fig. 2a, the spin-like and spin-unlike 1s IX with an opposite electric dipole have AMQNs of 0 and -1, respectively, neither of which matches that of $\mathrm{X_T}$ or $\mathrm{X_B}$. As a result, no brightening can happen to the 1s IX in the positive field direction, regardless of its spin configuration.

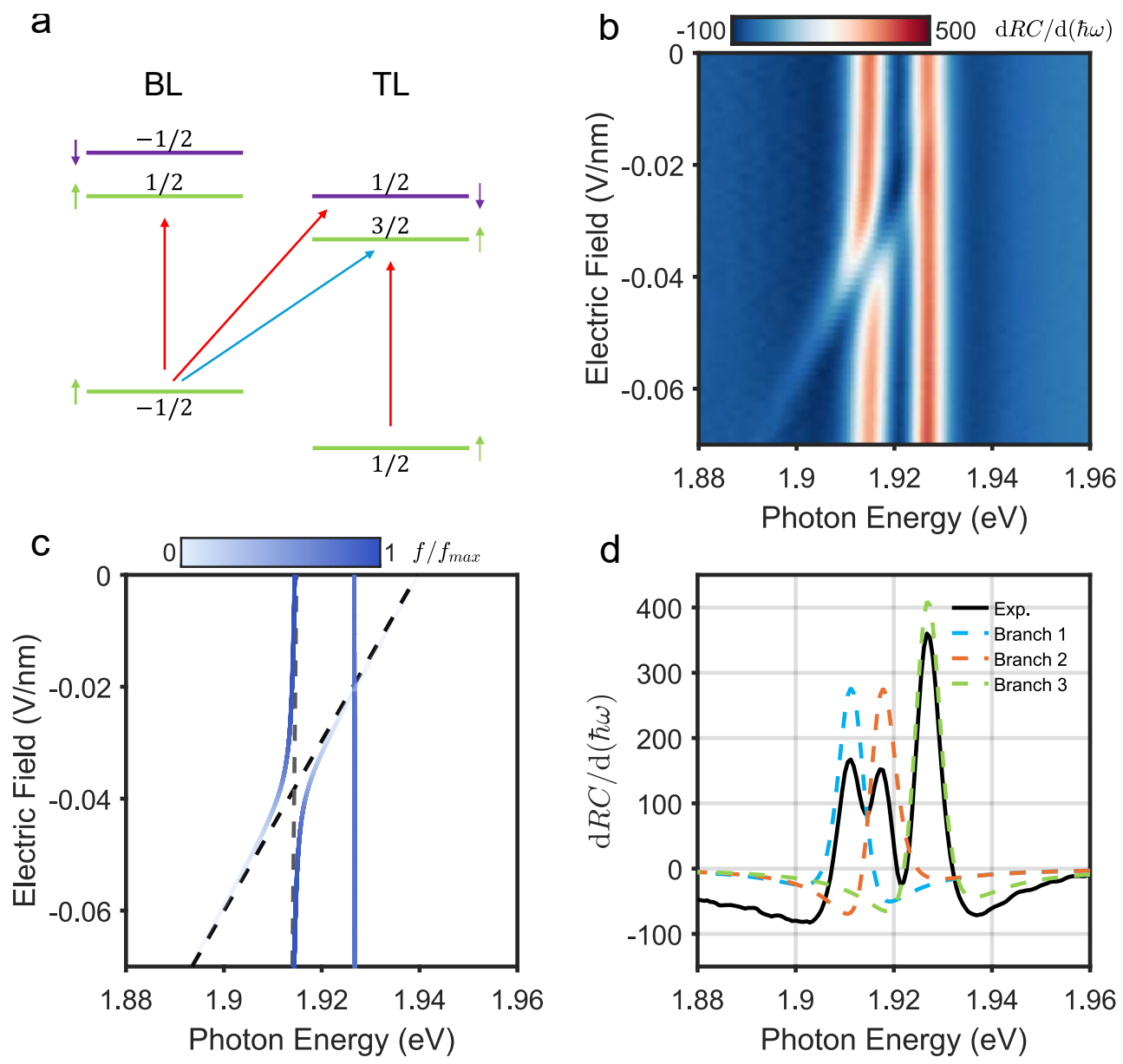

FIG. 2. Asymmetric hybridization of the spin-unlike interlayer exciton. (a) AMQNs of the spin- and layer-resolved Bloch states at the K valley. (b) Zoomed-in view of the dRC spectra near $\mathrm{IX_1}$. (c) Fitted peak energies of hybridized excitons as a function of the electric field. Black dashed curves denote the energies of unhybridized states. (d) dRC spectrum at $E = -0.038\, V/nm$. Solid curve is the measured spectrum, while dashed curves highlight the individual contribution of each hybridized exciton.

*2p interlayer exciton with odd parity*—While the asymmetric hybridization pattern of $\mathrm{IX_1}$ reveals its spin-unlike nature, the emergence of a second positive-dipole IX ($\mathrm{IX_2}$) at larger negative fields presents a distinct phenomenon: $\mathrm{IX_2}$ exhibits two similar avoided crossings with both $\mathrm{X_A}$ and $\mathrm{X_B}$ (Fig. 3b). Through the same three-level model fitting, we obtain two nearly identical coupling strengths, $|t_{2B}| = 2.7\, meV$ and $|t_{2T}| = 2.6\, meV$. The symmetry is also reflected in the nearly identical oscillator strengths of the first and third hybridized excitons when the electric field is set in-between the two avoided crossings (Fig. 3c). Such a symmetric hybridization rules out the possibility of $\mathrm{IX_2}$ being a high-order Rydberg state of the spin-unlike IXs. Through symmetry analysis combined with first-principles GW-BSE calculations, we conclude that $\mathrm{IX_2}$ is the 2p state of a spin-like IX.

In two-dimensional hydrogenic series, the $2p^{\pm}$ Rydberg state hosts an envelope function with an azimuthal phase gradient, which contributes an additional AMQN of $\pm 1$ [20,35,36]. This envelope angular momentum can compensate the valley one, aligning the total angular momentum of $\mathrm{IX_2}$ with the intralayer excitons. The spin-like 2p IX also explains the symmetric hybridization – since $\mathrm{IX_2}$ only shares the electron or hole with $\mathrm{X_T}$ or $\mathrm{X_B}$ respectively, the coupling strengths among them should be similar. This assignment is fully consistent with first-principles calculations and analyses. Fig. 3d shows the GW-BSE-calculated exciton envelope function of $IX_2$, which exhibits a ring-like shape with negligible amplitude at the zero electron-hole separation, a characteristic of the p series Rydberg state [2].

On the other hand, the hybridization between $IX_2$ and intralayer excitons exhibits an unprecedented interference effect that can resolve the phase of the constituent Bloch states. As shown in Fig. 3c, when $IX_2$ is field-tuned to the middle of two intralayer resonances, it acquires an oscillator strength comparable to the hybridized intralayer exciton. In contrast, once $IX_2$ falls below $X_B$ or rises above $X_T$, its oscillator strength diminishes rapidly, drastically different from a common avoided crossing behavior (e.g. between $IX_1$ and $X_B$). Here we attribute this unusual hybridization pattern to the quantum interference between transition dipoles within the hybridized $IX_2$ state.

In the phenomenological three-level model described above, the effective coupling between $IX_2$ and $X_T$ ($X_B$) is represented by $t_{2T}$ ($t_{2B}$). $t_{2B}$ and $t_{2T}$ have a similar amplitude, as indicated by the similar gap sizes in avoided crossings, but they do not necessarily have the same sign. In Fig. 3e, we plot the simulated dRC spectra by assuming $t_{2B}$ and $t_{2T}$ have either the same or opposite sign. Interestingly, if the sign is the same ($t_{2T}t_{2B} > 0$), the middle branch becomes strongly suppressed when $IX_2$ is centered between $X_T$ and $X_B$, which is clearly different from our experimental observations. The phenomenological model suggests the suppression results from coherent transition-dipole cancellation between the $X_T$ and $X_B$ components within the hybridized $IX_2$ state, an EIT-like destructive interference (Supplementary Information [41]). In contrast, oppositely signed couplings ($t_{2T}t_{2B} < 0$) reverse the relative phase between two dipoles, causing constructive interference and reproducing the experimentally observed enhancement of the middle branch (Fig. 3e and Supplementary Fig. S1 [41]). These analyses suggest that the $IX_2$ has an opposite coupling with intralayer excitons in different layers.

Microscopically, the opposite coupling can be explained by the composition and symmetry of the $IX_2$ wavefunction. Below we outline the physics underlying the phenomena while the detailed derivation can be found in Supplementary Information [41]. As a 2p exciton with a ring-like envelope function, $IX_2$ consists of electronic Bloch states at finite q away from the K point. In the two-band $\boldsymbol{q} \cdot \boldsymbol{p}$ model, the finite-q periodic part of Bloch states in the conduction/valence band of $l$-th layer is a mixture of states at the band edges: $|u_{c,l}(\boldsymbol{q})\rangle = |u_{c,l}(0)\rangle + \frac{\gamma q^+}{\Delta}|u_{v,l}(0)\rangle$ and $|u_{v,l}(\boldsymbol{q})\rangle = |u_{v,l}(0)\rangle - \frac{\gamma q^-}{\Delta}|u_{c,l}(0)\rangle$, where $\gamma$ is the interband coupling constant, $q^{\pm} = q_x \pm iq_y$, and $\Delta$ is the bandgap energy [20,33]. In Fig. 3f, we illustrate this mixing picture utilizing the distinct d-orbital character at the band extrema in TMDs.

Based on the GW-BSE theory, the exciton coupling constant $t$ is proportional to the product of the wavefunction overlap integrals between electrons and between holes. Since $IX_2$ shares the electron (hole) band with $X_T$ ($X_B$), one of the two overlap integrals is nearly unity. To evaluate the other integral, we note that, at the exact K point ($q = 0$), the valley angular momentum forces all overlaps to vanish except for the interband overlap between the conduction band in the bottom layer and the valence band in the top layer [19,24], $S = \langle u_{c,B}(0) | u_{v,T}(0) \rangle$. As a result, the overlap integrals associated with the unshared electron and hole are given by $\langle u_{c,B}(0) | u_{c,T}(\boldsymbol{q}) \rangle \propto +\frac{\gamma S}{\Delta}$ and $\langle u^{*}_{v,T}(0) | u^{*}_{v,B}(\boldsymbol{q}) \rangle \propto -\frac{\gamma S}{\Delta}$, respectively. Clearly, the two integrals are equal in magnitude but opposite in sign, giving rise to the opposite coupling constants as observed in experiments. Such a sign contrast is gauge invariant, as guaranteed by the orthogonality of the conduction- and valence-band Bloch spinors.

Regarding the negative-dipole counterparts of 2p IXs, the additional envelope AMQN allows a corresponding branch ($2p^{\pm}$) in either spin-like or spin-unlike configuration to hybridize with intralayer excitons. These negative-dipole IXs were not observed because the band offset shown in Fig. 1b places the negative-dipole IXs approximately 130 meV above their positive-dipole counterparts [24]. Tuning them into resonance with the intralayer excitons requires electric fields beyond the accessible range in our experiments.

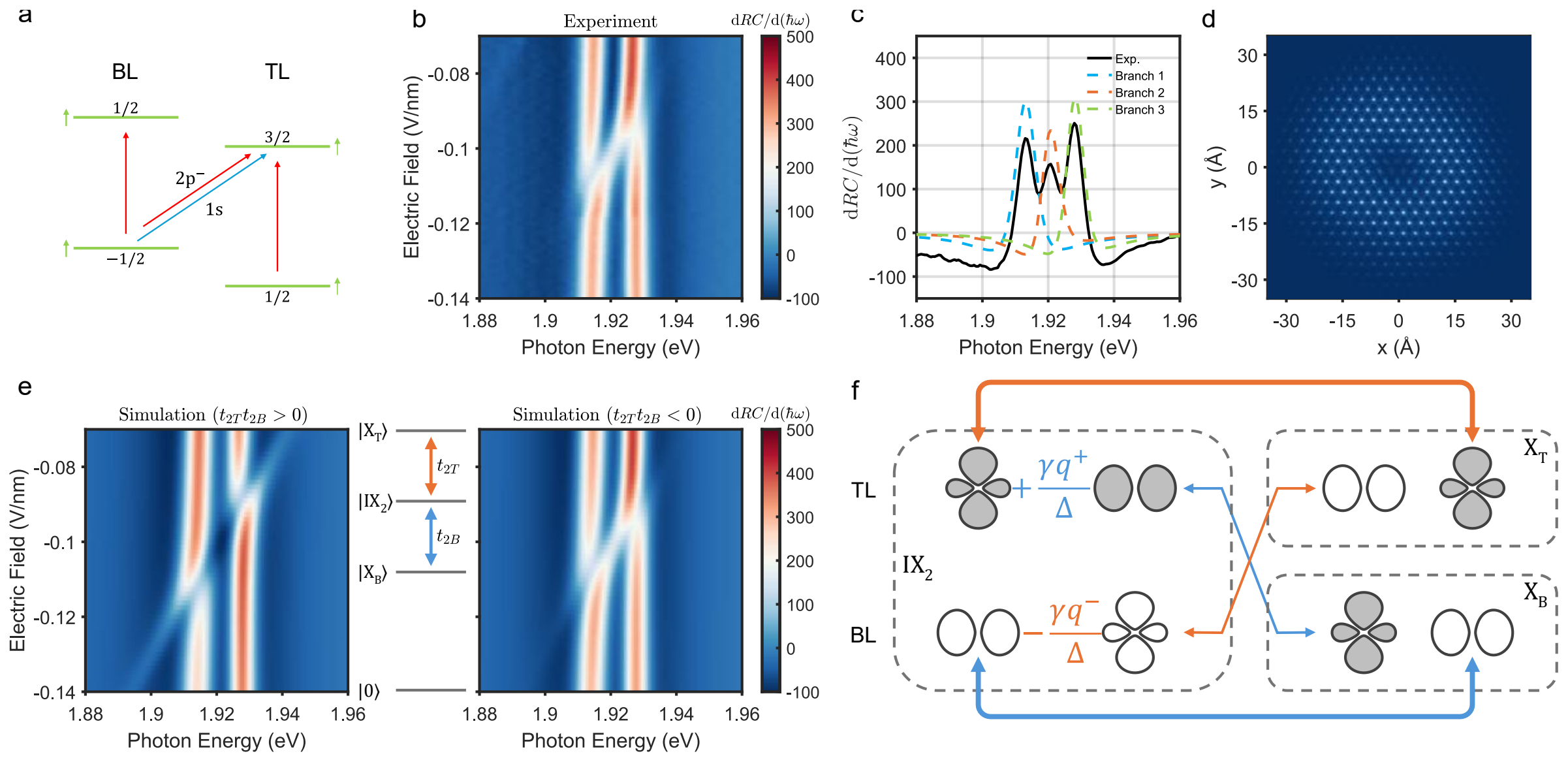

FIG. 3. Symmetric hybridization and quantum interference effect of a parity-odd 2p interlayer exciton. (a) Schematic of the $2p^{-}$ IX, whose AMQN is changed to +1 due to the envelope function. (b) Zoomed-in view of dRC spectra near $IX_2$. (c) dRC spectrum at $E = -0.101\ V/nm$, with the contribution of each hybridized exciton

resolved. (d) Calculated real-space exciton envelope function of $IX_2$ projected onto the $MoS_2$ plane, with the hole position fixed near a Mo atom at the center of the plot. (e) Simulated dRC spectra for coupling constants of the same or opposite signs. The schematic in the middle illustrates the bare exciton states and the coupling between them. (f) Schematic of the microscopic origin of the coupling sign contrast. The filled and open symbols represent the $d_{z^2}$ and $d_{x^2-y^2} + id_{xy}$ characters of electrons and holes. Arrows indicate the wavefunction overlap: thick ones mean a near-unity overlap, whereas thin ones denote the small but finite wavefunction overlap arising from the interband mixing at finite q.

To further support the IX assignment, we perform helicity-resolved RC measurements under an out-of-plane magnetic field ($B = 9\,T$). The resulting Zeeman splitting, $\Delta E = E_{\sigma^+} - E_{\sigma^-} = g\mu_B B$, was extracted for both the intralayer excitons and IXs. In TMDs, the magnetic-field-induced shift of a band-edge state arises from the magnetic moments of spin, atomic orbital, and valley angular momentum associated with the Berry curvature [32,33,45,46]. Depending on its specific configuration, an exciton can exhibit a Zeeman splitting comprising some or all of them. Experimentally, we found that $IX_1$ at moderate detuning has a g-factor of $-9.4 \pm 0.3$, a value larger than that of bright excitons ($-4.5 \pm 0.2$) by about 4, consistent with the expectation for the additional spin contribution (Fig. 4a). In the meantime, $IX_2$ also exhibits an enhanced g-factor compared to the bright excitons ($g = -5.7 \pm 0.3$), agreeing with the recent theory about the Zeeman splitting of a non-s Rydberg state [47], although the detailed calculation is beyond the scope of this work. The observed g-factor enhancements thus corroborate with our assignments of $IX_1$ and $IX_2$.

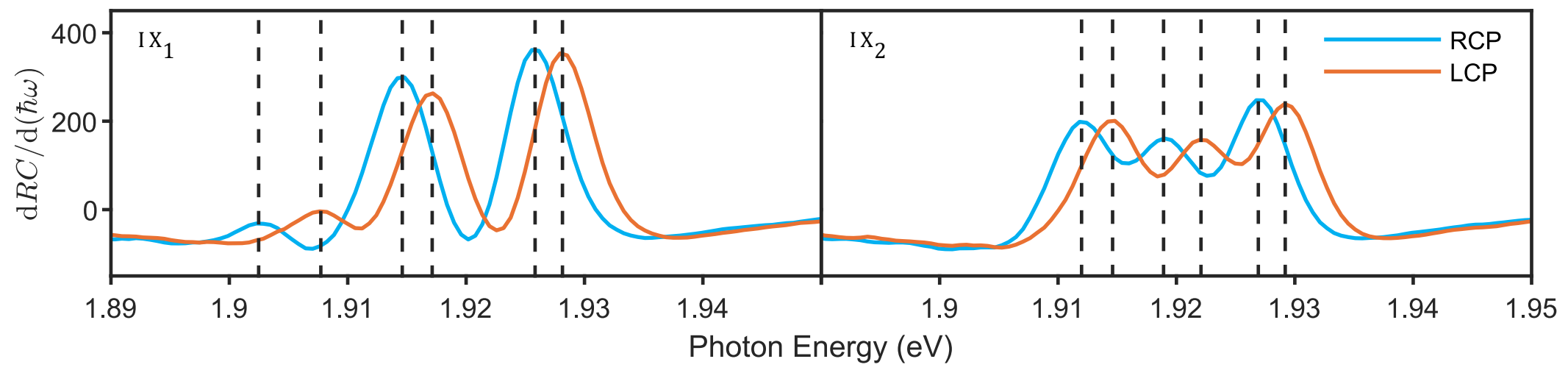

Fig. 4. Helicity-resolved dRC spectra, measured under an out-of-plane magnetic field of $B = 9\,T$ at electric fields $E = -0.049\,V/nm$ ($IX_1$) and $E = -0.101\,V/nm$ ($IX_2$). Blue and orange curves denote left- and right-handed circularly polarized spectra (LCP and RCP). The vertical dashed lines highlight the resonance of each hybridized exciton and the enhanced Zeeman splitting of the spin-unlike 1s and parity-odd 2p interlayer excitons.

Finally, we estimate the IX binding energy by examining the resonance energy relationship between IXs and $X_B$. Based on the experimentally measured 1s-2s splitting of about 156 meV for the intralayer exciton in our sample, we expect $X_B$ to have a total binding energy ranging from 176 meV (2D hydrogen model) to 203 meV (screened Keldysh potential [48]). At zero external electric field, $IX_1$ lies about 25 meV above $X_B$, due to its smaller single-particle gap and the binding energy difference. Having holes in the same valence band and electrons in conduction bands of opposite layers, the single-particle gap difference is mainly set by the 58-meV layer potential difference [24] plus a small conduction band spin splitting. Therefore, we deduce the binding energy of $IX_1$ to be about 120 meV. The IX total binding energy is smaller than that of intralayer excitons, due to the larger spatial separation between electron and hole.

In summary, we observe the brightening of two symmetry-forbidden interlayer excitons in 3R-$MoS_2$ bilayers with unique hybridization patterns that reveal a hidden phase contrast between the underlying Bloch states and might find applications in optical spin injection.

*Acknowledgement*—This work was supported by the Natural Sciences and Engineering Research Council of Canada (NSERC ALLRP/578466-2022 and RGPIN-2025-06880), the Canada Foundation for Innovation, the Canada First Research Excellence Fund, the Max Planck-UBC-UTokyo Center for Quantum Materials, the Gordon and Betty Moore Foundation's EPiQS Initiative (GBMF11071), and Canada Research Chair program (Z.Y.). The theoretical work is mainly supported by the United States DOE, Office of Basic Energy Sciences, under the Contract No. DE-SC0026336 (Z.L.). C.E.H. and H.C.H. acknowledge the support by the National Science and Technology Council (NSTC) in Taiwan under Grants NSTC 114-2112-M-032-009-MY3 and 114-2811-M-032-012-MY2. C.E.H is also supported by the funding from the University of Southern California. An award of computer time was provided by the INCITE program. This research used resources of both the Argonne and Oak Ridge Leadership Computing Facilities, which are DOE Office of Science User Facilities supported under contracts DE-AC02-06CH11357 and DE-AC05-00OR22725. Computational resources were also provided by National Energy Research Scientific Computing Center, which is a DOE SC User Facilities supported under contract DE-AC0205CH11231, and by Texas Advanced Computing Center, which is supported by United States NSF under Grant No. OAC-1818253.

Z.Y. conceived the project. H.H. fabricated the devices and performed all the optical measurements. C.E.H. performed the GW-BSE calculation under the supervision of H.C.H. and Z.L. K.W. and T.T. provided the hBN crystal. H.H., C.E.H., Z.L. and Z.Y. analyzed the data. H.H. and Z.L. proposed the phenomenological model. Z.Y. and H.H. wrote the manuscript based on the input from all other authors.

*Data availability*—The data that support the findings of this article are openly available [49].